\documentclass[11pt, journal, compsoc, onecolumn, a4paper]{IEEEtran} 

\usepackage[utf8]{inputenc} 

\usepackage{geometry} 
\usepackage{graphicx} 
\usepackage{hyperref}
\usepackage{amsmath}
\usepackage{color}
\usepackage{caption}

\usepackage{booktabs} 
\usepackage{array} 
\usepackage{paralist} 
\usepackage{verbatim} 
\usepackage{subfig} 
\usepackage{pdfpages}
\usepackage{setspace}
\usepackage{tabularx,ragged2e,booktabs,caption}
\newcolumntype{C}[1]{>{\Centering}m{#1}}

\usepackage{fancyhdr} 
\usepackage[titles,subfigure]{tocloft} 

\usepackage{caption}

\title{Hierarchical modelling of species sensitivity distribution: development and application to the case of diatoms exposed to several herbicides}
\author{\IEEEauthorblockN{Guillaume Kon Kam King\IEEEauthorrefmark{1}, Floriane Larras\IEEEauthorrefmark{2}, Sandrine Charles\IEEEauthorrefmark{1}\IEEEauthorrefmark{3} and Marie Laure Delignette-Muller\IEEEauthorrefmark{1}\IEEEauthorrefmark{4}}\\
\vspace{1em}
\IEEEauthorblockA{\IEEEauthorrefmark{1}CNRS, UMR5558, Laboratoire de Biométrie et Biologie \'Evolutive\\F-69622 Villeurbanne, France}\\

\IEEEauthorblockA{\IEEEauthorrefmark{2}Institut National de la Recherche Agronomique\\
UMR 0042, Carrtel\\
Thonon, France}\\
\IEEEauthorblockA{\IEEEauthorrefmark{3}Institut Universitaire de France\\ 103 bd Saint-Michel, 75005 Paris, France\\
Telephone: (800) 555--1212, Fax: (888) 555--1212}\\
\IEEEauthorblockA{\IEEEauthorrefmark{4}VetAgro Sup \\Campus Vétérinaire de Lyon, 69280 Marcy l’Étoile, France}\\
Corresponding author: guillaume.kon-kam-king@univ-lyon1.fr}

\begin{document}
\maketitle
\begin{abstract}The Species Sensitivity Distribution (SSD) is a key tool to assess the ecotoxicological threat of contaminant to biodiversity. It predicts safe concentrations for a contaminant in a community. Widely used, this approach suffers from several drawbacks: i)summarizing the sensitivity of each species by a single value entails a loss of valuable information about the other parameters characterizing the concentration-effect curves; ii)it does not propagate the uncertainty on the critical effect concentration into the SSD; iii)the hazardous concentration estimated with SSD only indicates the threat to biodiversity, without any insight about a global response of the community related to the measured endpoint. We revisited the current SSD approach to account for all the sources of variability and uncertainty into the prediction and to assess a global response for the community. For this purpose, we built a global hierarchical model including the concentration-response model together with the distribution law for the SSD. Working within a Bayesian framework, we were able to compute an SSD taking into account all the uncertainty from the original raw data. From model simulations, it is also possible to extract a quantitative indicator of a global response of the community to the contaminant. We applied this methodology to study the toxicity of 6 herbicides to benthic diatoms from Lake Geneva, measured from biomass reduction. 
\end{abstract}

\newpage

\tableofcontents

\section{Introduction}

\subsection{General introduction to SSD}

The Species Sensitivity Distribution (SSD) is a cornerstone in ecological risk assessment. 
Among other uses, it serves to predict concentrations in contaminant which are safe for a community. 
SSD is essentially an extrapolation of the sensitivity of a community from monospecific laboratory tests. 
The most standard approach\cite{Aldenberg1993,Aldenberg2000a,Posthuma2010} models the interspecific sensitivity variability in an assemblage of tested species in three steps.
In the first step, the sensitivity of each species is summarized by a single Critical Effect Concentration (CEC). 
This CEC can be a No Observed Effect Concentration (NOEC) or a Lowest Observed Effect Concentration (LOEC). It can also be a No Effect Concentration (NEC), or an Effective Concentration at x\% ($\mathrm{EC}_{x}$), which are obtained by fitting a model to the concentration-effect curve.
In the second step, the CECs in the community are assumed to follow a distribution law. Common choices for the distribution law include lognormal, loglogistic, BurrIII, \ldots The chosen distribution is then fitted to the CECs of the sample of tested species. 
In the third step, the Hazardous Concentration to p\% of the community ($\mathrm{HC_p}$) is computed as a percentile of the previous distribution.

The $\mathrm{HC_p}$ represents the concentration which is susceptible to \emph{affect} $\mathrm{p}\%$ of the community. The term ''\emph{affect}'' is directly linked to the type of CEC in terms of level of effect (for example the x of the $\mathrm{EC}_{x}$) and of biological effect (lethal, non-lethal, acute, chronic). With NOECs or NECs, one expects the $\mathrm{HC_p}$ to leave $(100-\mathrm{p})\%$ of the community species completely unharmed. Using $\mathrm{EC_{50}}$ however, which is a level of effect commonly selected, one expects  $(100-\mathrm{p})\%$ of the community to remain unaffected, which means that they suffer a reduction of less than 50\% to their measured endpoint. But it is not possible to determine the reduction suffered by the unaffected species, which could lie anywhere between 0 and 50 \%. 

SSD essentially carries information about the \emph{structural} response of a community to a contaminant, ie. the fraction of species affected at a certain level. The $\mathrm{HC_p}$ for small $p$, such as the  $\mathrm{HC_5}$, is ultimately used as a risk indicator. It is compared to the actual concentration of contaminant in an environmental setting to determine if the community living there is at risk, or to define an acceptably safe concentration for that community.

Several sources of uncertainty enter at the various steps of the SSD approach and have an influence on the predicted $\mathrm{HC_p}$ value. Firstly, there is an uncertainty on the estimate of the CEC from the experimental data: when the CEC is estimated from a concentration-effect curve or more generally from any model, it comes with a confidence interval.
Secondly, uncertainty arises from the fitting of a distribution to the CECs:  parameters of the distribution also have their own confidence intervals. This adds to the total uncertainty on the HC5. The uncertainty of this second step has already been studied and methods have been found for specific distribution laws\cite{Aldenberg1993,Aldenberg2000a,Wagner1991}. For other types of distributions, it is possible to use bootstrap\cite{efron1993introduction} to obtain confidence intervals, as described by Shao for the BurrIII distribution\cite{Shao2000} or in previous work by the authors\cite{ETC:ETC2644}. This uncertainty was also investigated with non parametric approaches in the estimation of the SSD\cite{Jagoe1997,Verdonck2001,VanderHoeven2001,Grist2009a}. However, there are currently very few attempts to include at the same time all the sources of uncertainty into the final prediction of the SSD\cite{Aldenberg2013}. 

\subsection{Several flaws of current SSD methodology}

The classical SSD approach described in the previous section and its many variants present a number of flaws\cite{Forbes2002,doi:10.1021/es972418b} ranging from ecotoxicological concerns (use of laboratory data to predict field effects, inferring community sensitivity from monospecific sensitivities, chronic vs. acute effects \ldots) to statistical issues (fitting a distribution on a small dataset, distributional assumptions, treatment of the uncertainty, \ldots).

This paper focuses on several of these: 
first, the classical SSD approach does not propagate the uncertainty on the CEC to the prediction. This is a source of concern, because following this approach, the uncertainty on the $\mathrm{HC_p}$ depends on the number of species, but not on the quality of the data used. 
Second, the CEC retains only a fraction of the information originally present in the data. Since the aim of SSD is to model the variability in sensitivity in the community, it is important to consider all the information available in the data. Indeed, there is relevant biological information in the parameters of the concentration-effect curve and their potential correlations.
Third, providing an $\mathrm{HC_p}$, the classical SSD approach outputs information about a structural response of the community only. It essentially yields the proportion of affected species for a given concentration in contaminant. It does not give information about the global response of the community\cite{Forbes2002,Kefford2012,DeLaender2008a}, ie. a response of the same nature as the measured endpoint. For instance, when using $\mathrm{EC_{50}}$ for biomass reduction as input, the SSD does not say anything about the change in the biomass of the community. In other words, the SSD aims to protect the structure of the community, but does not consider the effect on the community endpoint linked to the tested species which could be growth, reproduction, biomass, respiration, photosynthesis or any ecosystem process.


To address such issues, we revisited the current SSD approach to account for all the sources of variability and uncertainty into the prediction and to assess the risk for the community from a global point of view. For this purpose, we built a hierarchical model inspired by \cite{Moore2010} including the concentration-effect model together with the distribution law of the SSD.
From this hierarchical model, we were able to develop : 1) an indicator for the global response of the community, which we compared to the structural response predicted by the classical SSD; and 2) an SSD calculated on any level of effect ($x$ of the $\mathrm{EC}_x$) including interspecies correlation and all the uncertainty from the original data.

%

\section{Materials and methods}

\subsection{Diatoms dataset}

Our work was developed on a previously published dataset\cite{Larras2012} containing 10 diatom species exposed to 6 herbicides : atrazine, terbutryne, diuron, isoproturon, metolachlor and dimetachlor. Diatoms were unicellular microalgae which form a group of high diversity. The selected species of diatoms are representative of their community, and covered a great diversity in terms of taxonomy, morphology, sensitivity and ecological traits. Diatoms are often used to monitor water quality. The sensitivity of the species was determined assessing the growth  over four days as endpoint, based on chlorophyll a fluorescence, a proxy of the biomass. Bioassays were conducted in triplicates, on diatom strains in the exponential growth phase, when the daily growth ratio is approximately constant. Seven to ten herbicide concentrations were tested. Chlorophyll a fluorescence was measured using Fluoroskan (Fluoroskan Ascent, Thermo-scientific, Finland) at the beginning and at the end of the experiment. 


%


\subsection{Concentration-effect model}

Contrasting with \cite{Larras2012}, the response of each set (contaminant, species, replicate) was defined as the ratio:

\begin{equation}
\mathrm{R=\frac{\beta_{f}}{\beta_{0}}}
\end{equation}
where $R$ is the response, $\beta_{f}$ the fluorescence after 4 days and $\beta_{0}$ the initial fluorescence. Taking the logarithm of R and dividing by four, it would represent the daily exponential fluorescence growth rate, a proxy for the daily biomass exponential growth rate. Dividing by four has no influence on the results as we consider the relative reduction in fluorescence. Thus, we chose to ignore it for the sake of simplicity. Given the small number of replicates and given that this is not the focus of this article, we also chose not to model the replicate-effect, essentially grouping the three replicates together.

The response of a species $j$ to a given herbicide at concentration $x_i$ was modelled using a three parameter loglogistic model:

\begin{equation}
\mathrm{R=\frac{d}{1+\left(\frac{C}{e}\right)^{b}}}\label{eqR}
\end{equation}
where $C$ stands for the concentration,  $d$ is the response in the control, parameter $e$ is the $\mathrm{EC_{50}}$ in this model, ie. the concentration which induces a reduction of $50\%$ with regards to the response in the control. $b$ is a shape parameter, which is proportional to the slope of the concentration-effect curve at concentration $x=e$. By extension, parameter $b$ is usually called the slope of the concentration-effect curve, although the real slope at $x=e$ is $-\frac{d}{4e}b$.

A log-transformation of the data was necessary to avoid heteroscedasticity when estimating model parameters. Therefore, the following error model was used: 

\begin{equation}
\mathrm{Y=\ln(R)}+\epsilon=\mathrm{\ln\left(\frac{d}{1+\left(\frac{C}{e}\right)^b}\right)}+\epsilon \label{eqmodel}
\end{equation}
where $Y$ is the natural logarithm ($\ln$) of the measured endpoint, and $\epsilon \sim \mathcal{N}\left(0,\sigma \right)$

Parameter $d$ was estimated as the mean of the response in the control replicates for all the herbicides.
Parameters $b$ and $e$ were estimated by fitting model \ref{eqmodel} to observed data at the other concentrations to avoid using data twice. 
We chose to estimate parameter $d$ separately, because we were not interested in modelling or predicting the response in the control experiment. Only parameters $b$ and $e$ characterise the effect of the herbicide on the diatom.

%
%

\subsection{Classical SSD}

For each concentration-effect curve, we first fitted the model from eq.(\ref{eqmodel}) by non linear regression using the R package \emph{nlstools}\cite{Baty2013} and extracted the $\mathrm{EC_{10}}$ and the $\mathrm{EC_{50}}$ in order to compare two levels of effect. We computed bootstrap $95\%$ confidence intervals using non-parametric bootstrapping. Confidence intervals were also computed using the delta method\cite{weisberg2005applied}, and were found to be similar in general. Preference was given to the bootstrap method, because it prevents the lower bound of the interval from being negative, whereas the delta method sometimes gave negative values. 
Then, we fitted two lognormal distributions to the set of $\mathrm{EC_{10}}$ and $\mathrm{EC_{50}}$ using maximum likelihood via the web-tool MOSAIC\_SSD\cite{ETC:ETC2644} and obtained the $\mathrm{HC_{5}}$ for the community. 


\subsection{Hierarchical species sensitivity distribution}

\subsubsection{Hierarchical approach}

A hierarchical approach is very different from the fitting of individual concentration-effect curves. 
The philosophy behind the hierarchy is that all tested species represent a random sample from the community, and that their responses follow a distribution. More precisely, parameters $b$ and $e$ of the concentration-effect model are assumed to follow a multivariate distribution. This reasoning falls in line with the classical SSD, where CECs of the species are assumed to be sampled from a community sensitivity distribution. 
However, in the hierarchical approach, the whole response is assumed to be a sample from the community. 
This allows us to reconstruct the response of the whole community.
Another difference with the classical SSD approach is that the parameters of the community, called the hyperparameters, are estimated in one stroke from all the experimental data.
 This provides the advantage of pooling all the information together. Species for which the data are of very good quality will have the most important contribution to the global fit. Species for which the response is not characterized very precisely (large uncertainty on the parameters), or where data are missing, contribute less. In other words, all the data contribute to the estimation of the parameters at the extent of the information they contain. 
The classical SSD approach, on the contrary, heavily relies on the quality of the CEC estimates, and the requirements may be severe\cite{Dowse2013}. In the previous study of the diatom dataset, this entailed discarding all the data which did not allow to fit a concentration-effect model\cite{Larras2012}. 

\begin{figure}
\includegraphics{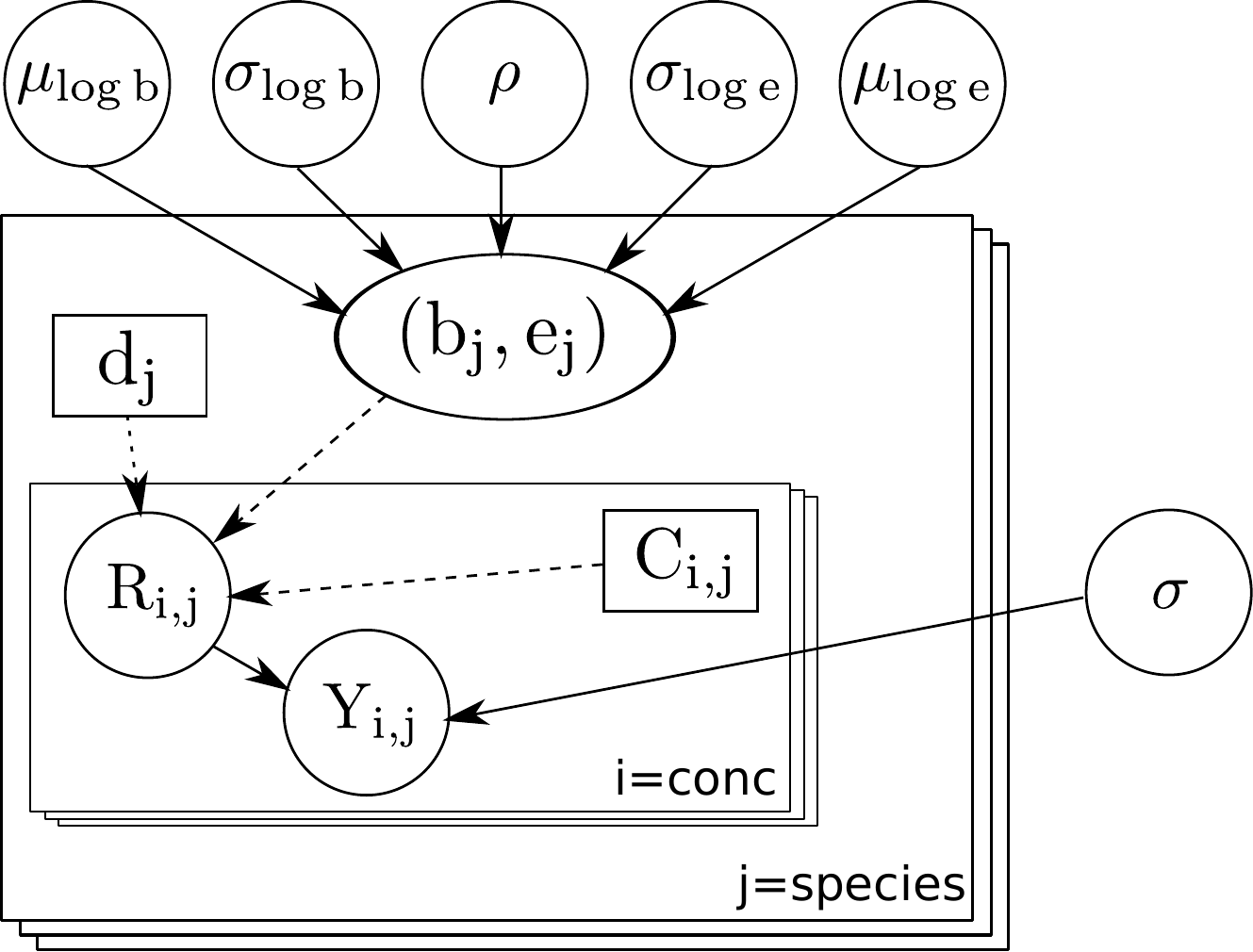}

\caption{Directed Acyclic Graph (DAG) of the hierarchical model and stochastic links in the model. Stochastic links are in solid lines, deterministic links in dotted lines. \label{DAG}}

\end{figure}

Fig. \ref{DAG} sketches the hierarchy in the model and table \ref{links} describes the links of the model.
Parameter $d$ having been estimated separately, we modelled the joint distribution of parameters $b$ and $e$. 
Both of them were assumed to follow a lognormal distribution. The lognormal distribution is the most commonly used for parameter $e$\cite{Wheeler2002} (which corresponds to the $\mathrm{EC}_{50}$). We assumed the same distribution for the $b$ parameter, knowing that the small number of species does not allow a better informed choice of distribution.
There could be a correlation between these two parameters, which we also modelled (parameter $\rho$). 
Therefore, $\log(b)$ and $\log(e)$ were assumed to follow a multivariate normal distribution ($\log$ is used for the base-10 logarithm).

\begin{table}
\caption{Description of the links indicated in Fig.\ref{DAG}} \label{links} 

\begin{center}
{\small
\begin{tabular}{lll}
\hline
  \textbf{Node} & \textbf{Type} &\textbf{Equation}\\
\hline
  $\mathsf{(\log\left(b_j\right),(\log\left(e_j\right))}$& Stochastic &$\mathsf{(\log\left(b_j\right),(\log\left(e_j\right))}\sim \mathcal{N}_m(\mathbf{\mu,\Sigma})$\\
  $\mathrm{R_{i,j}}$ & Deterministic & Eq. (\ref{eqR})\\
 $\mathrm{Y_{i,j}}$ & Stochastic &$\mathsf{Y_{i,j} \sim \mathcal{N}(\ln(R_{i,j}),\sigma)}$ \\
\hline
\end{tabular}
}

\end{center} where 
$\mathbf{\mu}= \left( \begin{array}{c}
 \mathrm{\mu_{\log b}} \\
 \mathrm{\mu_{\log e} }
 \end{array} \right)$, $\mathbf{\Sigma}= \left( \begin{array}{cc}
 \mathrm{\sigma_{\log b}^2} & \mathrm{\rho\sigma_{\log b}\sigma_{\log e}}  \\
 \mathrm{\rho\sigma_{\log b}\sigma_{\log e} }&  \mathrm{\sigma_{\log e}^2  }
 \end{array} \right)$, $ \mathcal{N}$ is the normal distribution and  $ \mathcal{N}_m$ is the multivariate normal distribution. 
\end{table}

%
%
The transition between the classical one-parameter SSD to our hierarchical model can be understood by the following  consideration: in classical SSD, one species is accounted for one value in the distribution and the whole community is represented by an univariate distribution. In our hierarchical model, each species is accounted for by a pair of values in a two dimensional distribution  and the whole community is represented by a multivariate distribution. 

%
%
%

%
%

%

\subsubsection{Bayesian methods}

We used JAGS\cite{plummer2003jags} to fit the hierarchical model. JAGS performs Bayesian inference using Gibbs sampling via Markov Chain Monte Carlo (MCMC) simulation. The priors are detailed in Table \ref{priors}. The prior on $\mu_e$ was a normal distribution centred on the middle of the range of all tested concentrations. Its standard deviation was defined so that $\mu_e$ had a 95\% probability to lie between the largest and the smallest tested concentrations. All the other priors were non informative. The chains were run for 500 000 iterations, and one in 40 were conserved.

\begin{table}
\caption{Prior distributions used for the hyperparameters of the hierarchical model (Fig.\ref{DAG})} \label{priors} 

\begin{center}
{\small
\begin{tabular}{clc}
\hline
  \textbf{Parameter} & \textbf{Distribution} &\textbf{Source of prior information}\\
\hline
 $\mathrm{ \mu_{\log b}}$&$\sim \mathcal{N}(-6,6)$& non informative\\
 $\mathrm{ \sigma_{\log b}}$ &$\sim \mathcal{N}(0,10)$ & non informative\\
  $\mathrm{\mu_{\log e}}$ &$\sim \mathcal{N}(\mathrm{\mu_{\log C}},\mathrm{\sigma_{\log C}})$& concentrations range\\
  $\mathrm{\sigma_{\log e}}$&$\sim \mathcal{U}(0,10)$& non informative\\
  $\rho $&$\sim \mathcal{U}(-1,1)$& non informative\\
  $\sigma $&$\sim \mathcal{U}(0,2)$& non informative\\
\hline
\end{tabular}
}
\end{center}
$\mathrm{\mu_{\log C}}=\frac{\log(min(C_{i,j}))+\log(max(C_{i,j}))}{2}$ and $\mathrm{\sigma_{\log C}}=\frac{\log(max(C_{i,j}))-\log(min(C_{i,j}))}{4}$. $\mathcal{N}(\mu,\sigma)$ denotes the normal gaussian distribution of mean $\mu$ and standard deviation $\sigma$, $\mathcal{U}(a,b)$ denotes the uniform distribution between $a$ and $b$.
\end{table}

The convergence of three chains was checked computing the Gelman-Rubin diagnostic\cite{Brooks1998}. Prior and posterior distributions were compared to check visually that the priors did not constrain the estimation of the posteriors.
The relative width of the prior and posterior distributions was also compared to ensure that sufficient information was learned from the data.  
The parameters of the hierarchical model came out as a joint posterior distribution. The median of the marginal distributions were used as estimates of the parameters. The $2.5$ and $97.5$ percentiles of the distribution were used to define a $95\%$ credible interval. An R script to fit the hierarchical model on the atrazine data is provided in the Supporting Information.


\subsection{Modelling the global response of a community}
Once the model fitted, the joint posterior distribution of the parameters contained all the information that can be extracted from the data about the response of the community to the contaminant. For a set of global parameters $\theta=(\mu_b,\sigma_b,\mu_e,\sigma_e,\rho)$ obtained from the posterior distribution, it was possible to reconstruct a full community by sampling individual species and to predict its response to the contaminant. Sampling a species $i$ is equivalent to sampling a pair of parameters $(b_i,e_i)$ from the multivariate normal distribution parametrised by $\theta$. 

In order to predict the response of a realistic community, we chose to focus on finite-size communities. Diatom communities may number around $30$ different species. Note that a specific draw of $30$ species produces a community with a certain response and that another draw of $30$ species would produce a different response. Therefore, there is some uncertainty in the response obtained for a group of 30 species, even assuming that $\theta$ is known. Moreover, the $\theta$ parameters themselves are uncertain and follow a distribution. These two sources of uncertainty were taken into account by sampling around 10 000 sets $\theta_k$, then sampling 30 species for each $\theta_k$. 


After a community was simulated, we defined its global response as the global fluorescence of the community, depending on the concentration. The global fluorescence was defined as the sum of the fluorescence of each species. 
To obtain a global response, we  assumed that all species participated equally in the global fluorescence. Following this assumption, it was possible to define an indicator of the global response of the community at a given concentration, called $\mathrm{r_{tot}}$:

\begin{equation}
\mathrm{r_{tot}}=\frac{\displaystyle{\sum_{i\in species}}\frac{\mathrm{R}_i}{\mathrm{R}_i^0}}{\mathrm{N_{species}}}
\end{equation}

where $\mathrm{R}_i$ is the response
of species $i$ at a given concentration, and $\mathrm{R}_i^0$ the response
in the control experiment. The indicator $\mathrm{r_{tot}}$ of the global response is a quantity between $0$ and $1$ which describes the global reduction in fluorescence growth compared to the control, as a function of the concentration in contaminant. Analogous to the $\mathrm{HC}_5$ for the SSD, a Global Effect Concentration of $5\%$ ($\mathrm{GEC}_5$) was defined, corresponding to the concentration leading to a reduction of $5\%$ of the global response  $\mathrm{r_{tot}}$.
 In our case, the $\mathrm{GEC}_5$ corresponds to a reduction by $5\%$ of the community fluorescence ($\mathrm{r_{tot}}=0.95$). Following the terminology used for SSD in Posthuma\cite{Posthuma2010}, the hierarchical SSD, and more precisely the prediction of the global response, can be used in a \emph{forward} and an \emph{inverse} manner. The \emph{forward} approach consists in setting a protective concentration threshold, the $\mathrm{GEC}_5$, below which $95\%$ of the global response of the community should be protected. The \emph{inverse} approach consist in determining the reduction in the global response of the community for a given concentration level.


\subsection{Hierarchical SSD with confidence intervals}

From the fitted model, it is also possible to reconstruct an SSD. In this case, the simulation aimed at representing the variability of species sensitivity, ie. the distribution of any $\mathrm{EC}_{x}$ in the community. For 2000 sets of global parameters  $\theta_k$, the concentration-effect responses (Eq. (\ref{eqR})) of a community of species were simulated, and their $\mathrm{EC}_{x}$ calculated. Estimating an $\mathrm{HC}_5$ for a community consists in determining the fifth percentile of that $\mathrm{EC}_{x}$ distribution. To get the best estimation of the fifth percentile, large communities were simulated ($4\times10^6$ species).
An SSD (and  $\mathrm{HC}_5$) with a 95\% credible interval was estimated for these 2000 sets of parameters, from the median, $2.5^{\text{th}}$ and $97.5^{\text{th}}$ percentiles. This  SSD is an improvement on the classical SSD, since it is estimated taking into account all the information present in the original data and accounting for the potential inter species correlation among the parameters $\mathrm{b}$ and $\mathrm{e}$ of the concentration-effect model. Especially, the uncertainty in the estimation of the parameters of the concentration-effect curves was propagated in the SSD estimation.
 Another advantage of reconstructing the SSD from the fitted hierarchical model is that it does not require to choose the $x$ of the $\mathrm{EC}_{x}$ in advance, contrary to the classical SSD, which starts with a certain level of  $\mathrm{EC}_{x}$. Fitting a classical SSD on another level requires going back to the original data. 
Using the fitted model and the simulation scheme instead, it is possible to calculate an SSD on any $x$ of the $\mathrm{EC}_{x}$. We used our hierarchical model to study how the $\mathrm{HC}_5$ may vary as a function of the $x$ of the $\mathrm{EC}_{x}$
\section{Results}

\subsection{Classical SSD}

\begin{figure}
\centering
\includegraphics[scale=.5]{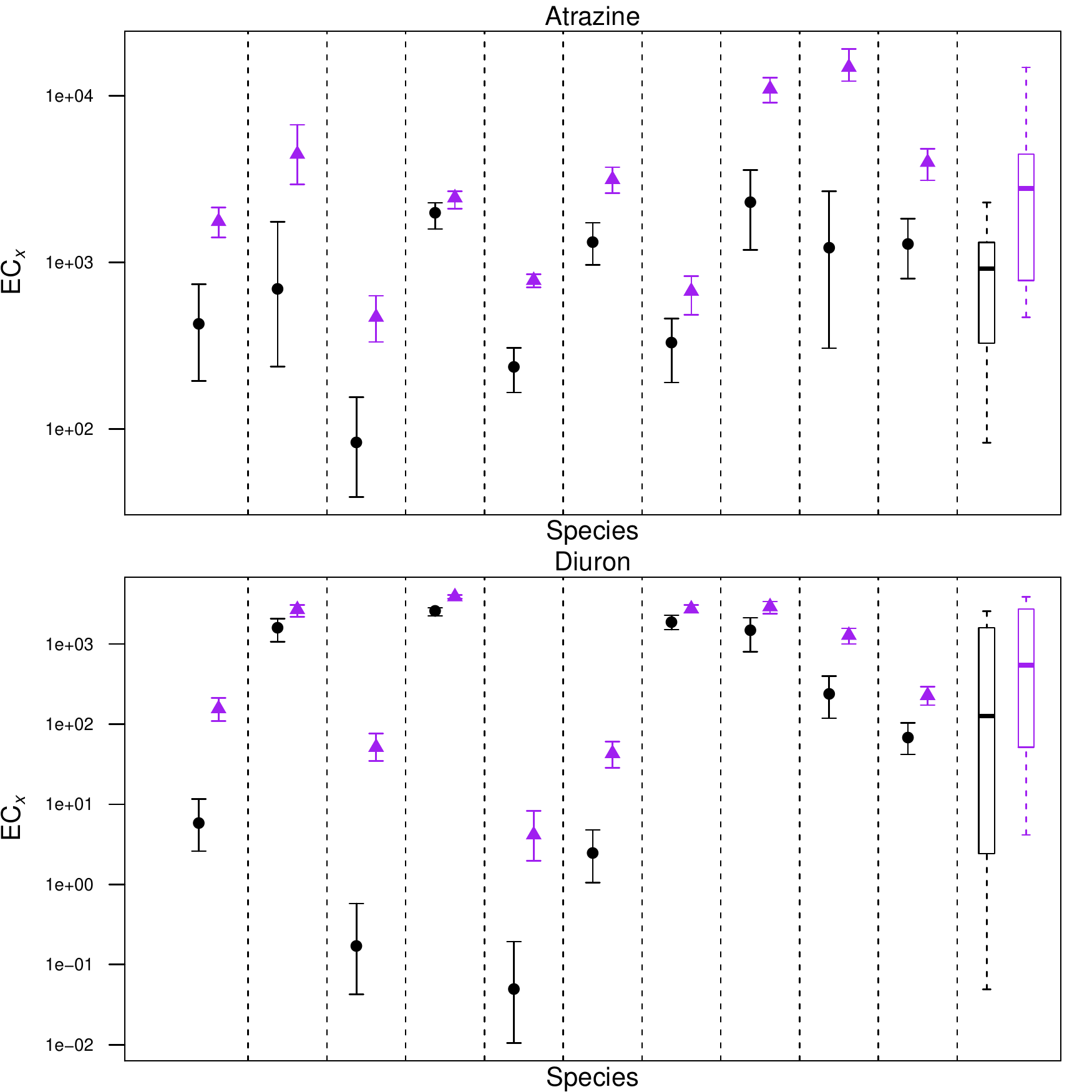}
\caption{$\mathrm{EC}_{10}$ in black and $\mathrm{EC}_{50}$ in blue for each species, with the $95\%$ bootstrap confidence interval. Vertical dotted lines separate each species. The box at the right hand side of each plot is the distribution of the point estimates of the $\mathrm{EC}_{x}$\label{fitted_param} at the corresponding level of $x$ for all the species.}

\end{figure}

The $\mathrm{EC}_{10}$ and the $\mathrm{EC}_{50}$ of each species were computed for every contaminants. Results for two contaminants are displayed on Fig. \ref{fitted_param}. For both herbicides, the confidence intervals appeared to be much larger on the $\mathrm{EC}_{10}$ than on the $\mathrm{EC}_{50}$.
Similar results were observed for the four other herbicides.

\subsection{Convergence of the MCMC algorithm}
The MCMC chains converged for all contaminants, according to the Gelman-Rubin statistics\cite{Brooks1998}. 
Fig. \ref{prior-post} shows for Diuron that except for parameter $\rho$, the non informative prior distributions did not constrain the posterior distribution of the parameters and that there was sufficient information in the experimental data to estimate them. Similar results were observed for the other herbicides. The apparent constraint on correlation parameter $\rho$ is natural, since the correlation lies between $0$ and $1$. The fit of the model was visualised  at the level of the original diatom species by superimposing the fitted curves on the original data.
The fitted curves were obtained by 
taking the median values of parameters $b_i$ and $e_i$ from the marginal posterior distributions. Fig. \ref{fit} shows that for atrazine and diuron, the estimation of the global parameters of the hierarchical model corresponds to a good fit at the level of the original diatom species. Results for the other herbicides can be found on Fig. \ref{fit} in the Supplementary Information.

\begin{center}
\includegraphics[width=0.7\textwidth]{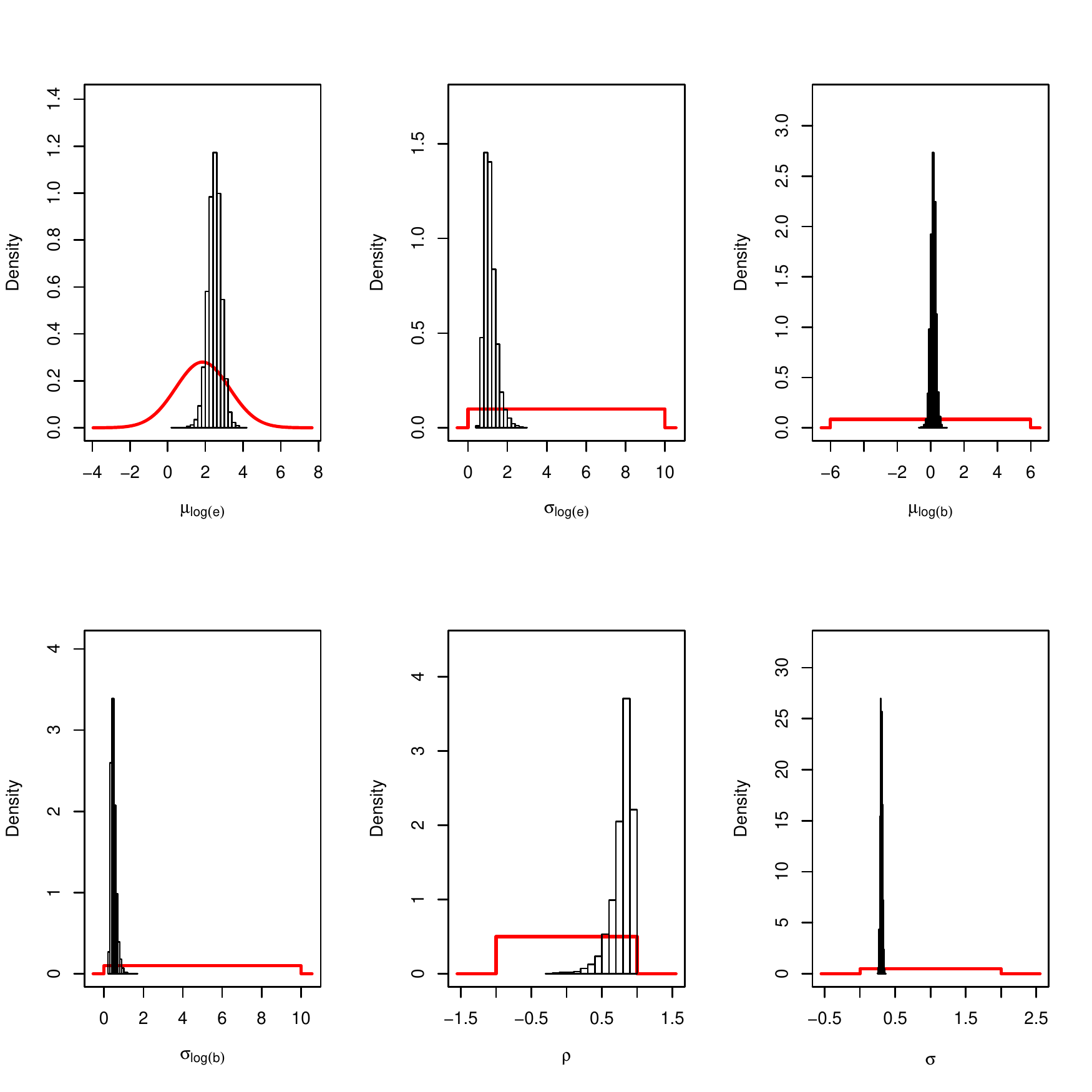}
\captionof{figure}{Comparison of the priors (in red) defined in Table \ref{priors} with the marginal posterior distributions (black histograms) for diuron. \label{prior-post}}

\includegraphics[width=0.4\textwidth]{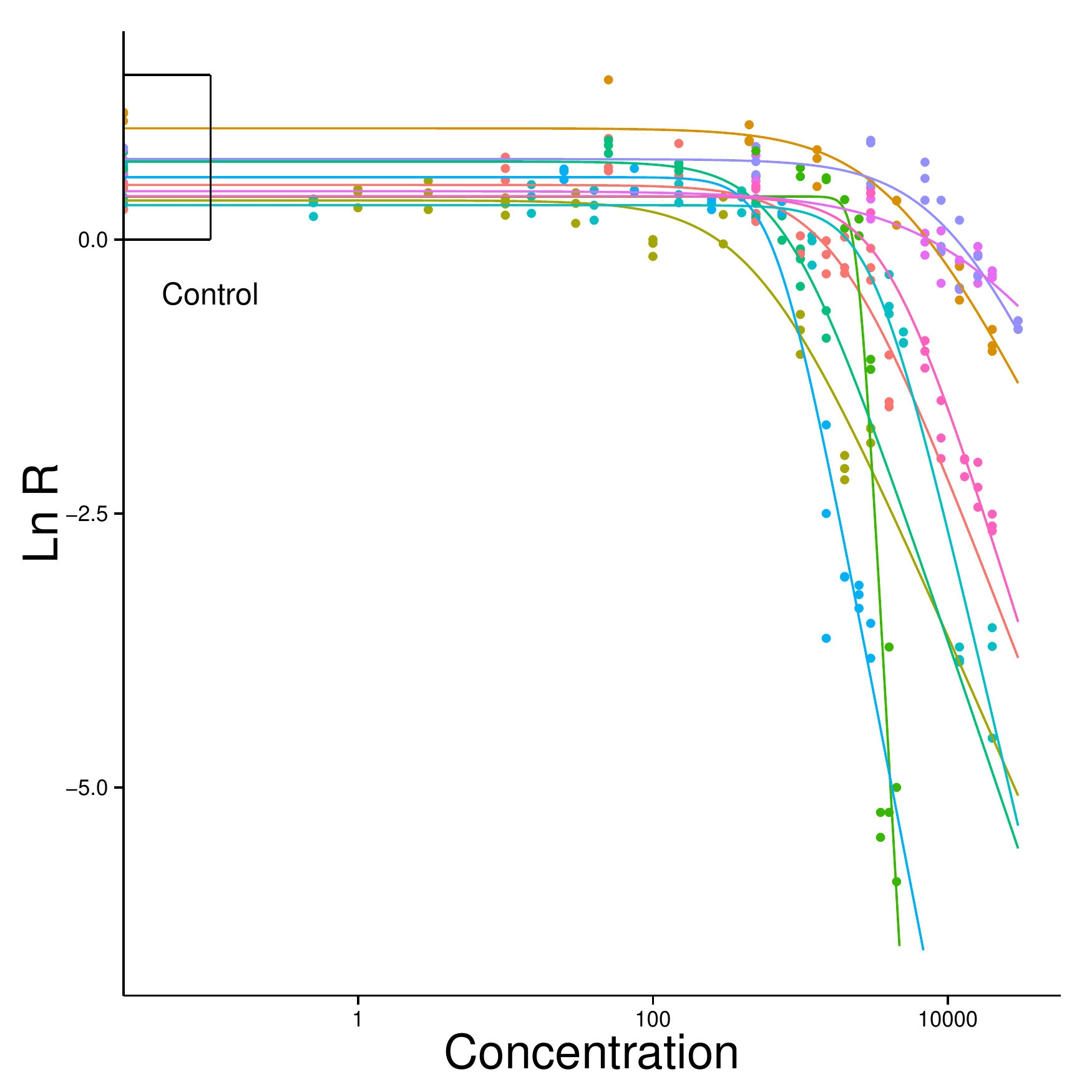}
\includegraphics[width=0.4\textwidth]{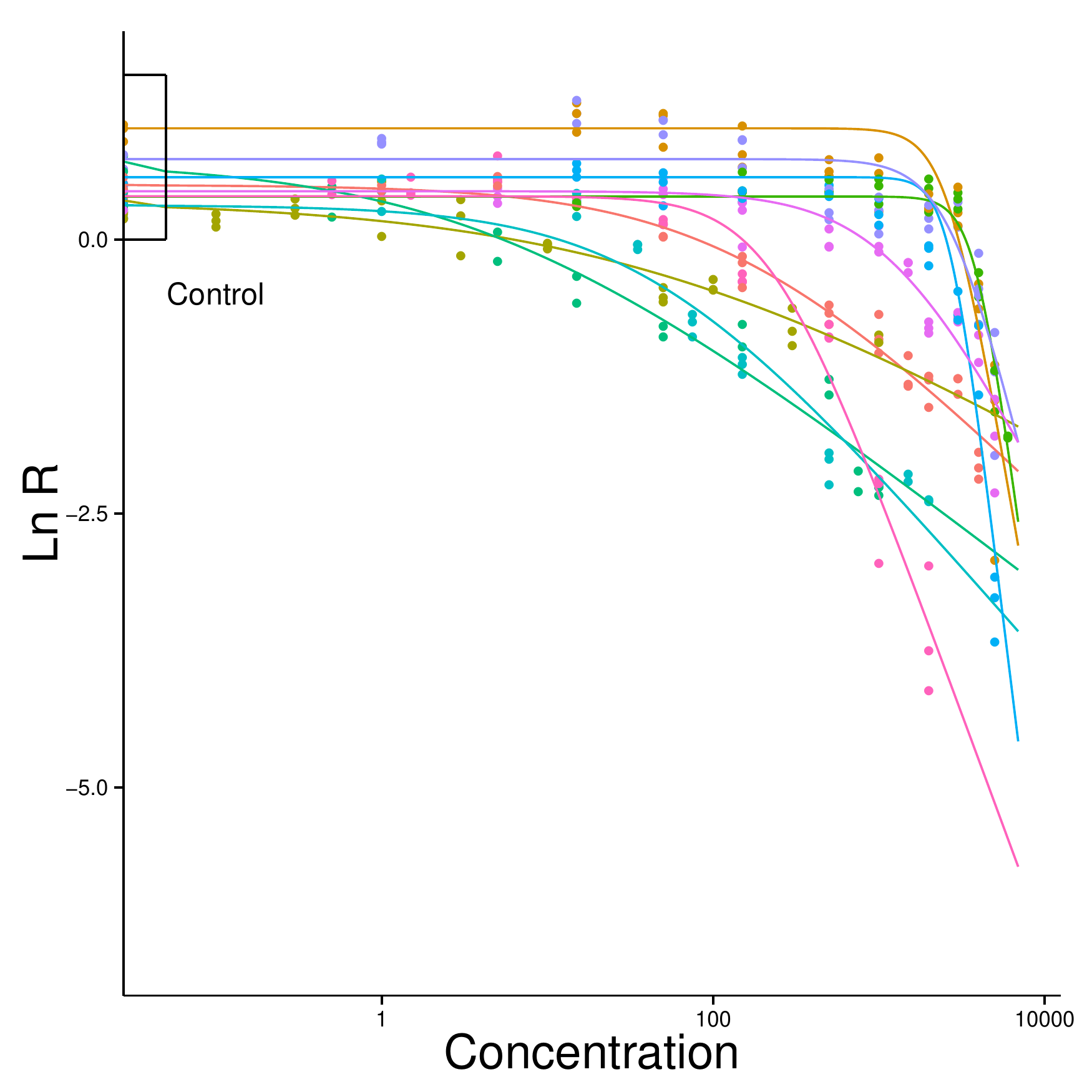}
\captionof{figure}{Original data and fit of the model at the level of the original diatom species for Atrazine (left) and Diuron (right). Each colour denotes a different species. The rectangle denotes the control response in log scale. Parameters for the curves are the median values of the marginal parameter distributions.\label{fit}}
\end{center}

\subsection{Parameter estimation}


The estimated parameters for atrazine and diuron are presented in Table \ref{param}, along with their $95\%$ credibility interval.  Table \ref{param} presents a summary of the marginal posterior distributions.
Table \ref{corr} contains the values of the correlation parameter for all herbicides. For atrazine, the $95\%$ credibility interval is centred around $0$, suggesting the absence of correlation between parameters $b$ and $e$. For all the other herbicides, however, there was a correlation between these two parameters. 
Slope parameter $b$ qualitatively determines how a species is affected by the contaminant: for a small value of $b$, the species is gradually affected by the contaminant, whereas for a large value of $b$, this species is almost insensitive to the contaminant up to a certain threshold, then suffers a drastic effect. 
A strong positive correlation between $b$ and $e$, the slope parameter and the $\mathrm{EC}_{50}$, implies that a species with a low slope parameter also has a small $\mathrm{EC}_{50}$, ie. the most sensitive species are affected gradually. A positive correlation also implies that the most resilient species show no effect up to a certain threshold, followed by a sudden drop in fluorescence. In the absence of correlation, there is no constraint on the relative value of $b$ and $e$ for a given species, and all types of behaviours can be encountered. The effects of that correlation are apparent on Fig. \ref{fit}: atrazine, a contaminant for which species show no correlation between parameters $b$ and $e$, shows all sorts of behaviours, for diuron, the species with small $\mathrm{EC}_{50}$ have a gradual slope and those with large $\mathrm{EC}_{50}$ have a steep slope.

Such information about correlation between the dose-response parameters is not considered or taken advantage of using the classical SSD approach. Yet this is an information of biological relevance which can only by addressed through the hierarchical modelling of SSD.\\


\begin{center}

\captionof{table}{Median parameters of the hierarchical model and their $95\%$ credibility interval, for atrazine and diuron. 
\label{param}}
\begin{tabular}{ccc}
\hline
   & Atrazine&Diuron\\

   \textbf{Parameter} & \textbf{Estimate}&\textbf{Estimate} \\
  \hline
$\log_{10} \mu_b$ & $0.28[0.02,0.55]$& $0.16[-0.15,0.46]$\\

$\log_{10} \sigma_b$ & $0.37[0.22,0.69]$& $0.46[0.30,0.82]$\\
  
 $\log_{10} \mu_e$& $3.36[2.93,3.75]$& $2.49[1.76,3.16]$\\
  
  $\log_{10} \sigma_e$&$0.58[0.37,1.12]$&$1.07[0.70,1.9]$\\
  
  $\rho$&$-0.22[-0.74,0.47]$&$0.83[0.39,0.96]$\\
  \hline
\end{tabular}

\captionof{table}{Correlation parameters for all herbicides and their $95\%$ credibility interval. \label{corr}}
\begin{tabular}{cc}
\hline
   \textbf{Pesticide}& $\rho$\\
  \hline
Atrazine &$-0.22[-0.74,0.47]$ \\

Terburtyne & $0.68[0.52,0.91]$\\

Diuron & $0.83[0.39,0.96]$\\
  
  Isoproturon &$0.87[0.78,0.97]$ \\
  
Metolachlor &$0.41[0.14,0.89]$ \\
  
Dimetachlor &$0.85[0.64,0.99]$\\
  \hline
\end{tabular}

\end{center}

%
%
%

\subsection{Modelling the global response of a community}

The hierarchical SSD approach extracts more information from the raw data than the classical SSD. 
It provides an indicator of the global response of the community, which is a relevant information for the protection of that community. 
Fig. \ref{HC_GEC_Diu} shows the importance of considering the global response of the community for risk assessment. 
The top of Fig. \ref{HC_GEC_Diu} shows the $\mathrm{HC}_5$ obtained using the classical SSD approach on the $\mathrm{EC}_{50}$ endpoint, while the bottom shows the $\mathrm{HC}_5$ obtained using the $\mathrm{EC}_{10}$. 
This concentration is used for regulatory purposes as the Predicted No Effect Concentration (PNEC), which determines the threshold under which the community is considered protected. 
The $\mathrm{HC}_5$ only aims at preventing a proportion of the species from being harmed, disregarding the possibility that harming key species could endanger the whole community. 
In order to protect the community in terms of the endpoint measured in the original data (fluorescence, biomass), it is interesting to consider also the $\mathrm{GEC}_5$ in the risk assessment. 
In the case of Atrazine, the concentration which induces a reduction of $5\%$ of the global fluorescence ($\mathrm{GEC}_5$) is lower than both the $\mathrm{HC}_5$ based on the $\mathrm{EC}_{50}$ and on the $\mathrm{EC}_{10}$ (see on Fig. \ref{HC_GEC_Diu}). 
In the case of Diuron, the $\mathrm{GEC}_5$ is much lower than the $\mathrm{HC}_5$ based on the $\mathrm{EC}_{50}$ and similar to the $\mathrm{HC}_5$ based on the $\mathrm{EC}_{10}$.  For the four other herbicides, the $\mathrm{GEC}_5$ is between the two $\mathrm{HC}_5$ and in general, the $\mathrm{GEC}_5$ is close to the $\mathrm{HC}_5$ based on EC10.
Calculating of the reduction in global fluorescence at the $\mathrm{HC}_5$ (ie. using the \emph{inverse} approach) indicates that for Atrazine, the classical $\mathrm{HC}_5$ built on the $\mathrm{EC}_{50}$, which protects $95\%$ of the species, could induce a reduction of 81\% [55\%,94\%] of the global fluorescence. 
The classical $\mathrm{HC}_5$ based on the $\mathrm{EC}_{10}$ could induce a reduction of 92\% [73\%,99\%] of the global fluorescence. 
In the case of Diuron, the classical  $\mathrm{HC}_5$ built on the  $\mathrm{EC}_{50}$ protects $86\% [69\%,96\%]$ of the fluorescence, while the classical $\mathrm{HC}_5$ built on the  $\mathrm{EC}_{10}$ protects $96\% [86\%,100\%]$ of the global fluorescence. 

To summarize the comparison, there is no systematic relationship between the $\mathrm{GEC}_5$ and the $\mathrm{HC}_5$. Aiming to protect $95\%$ of the global response of the community could prove either more or less protective than aiming to protect $95\%$ of the species. More precisely, using the forward approach we can estimate that for Atrazine and Diuron, a $\mathrm{HC}_5$ based on the $\mathrm{EC}_{50}$ might protect only $80-86\%$ of the global response of the community. 


\includegraphics[width=7.5cm,height=15cm]{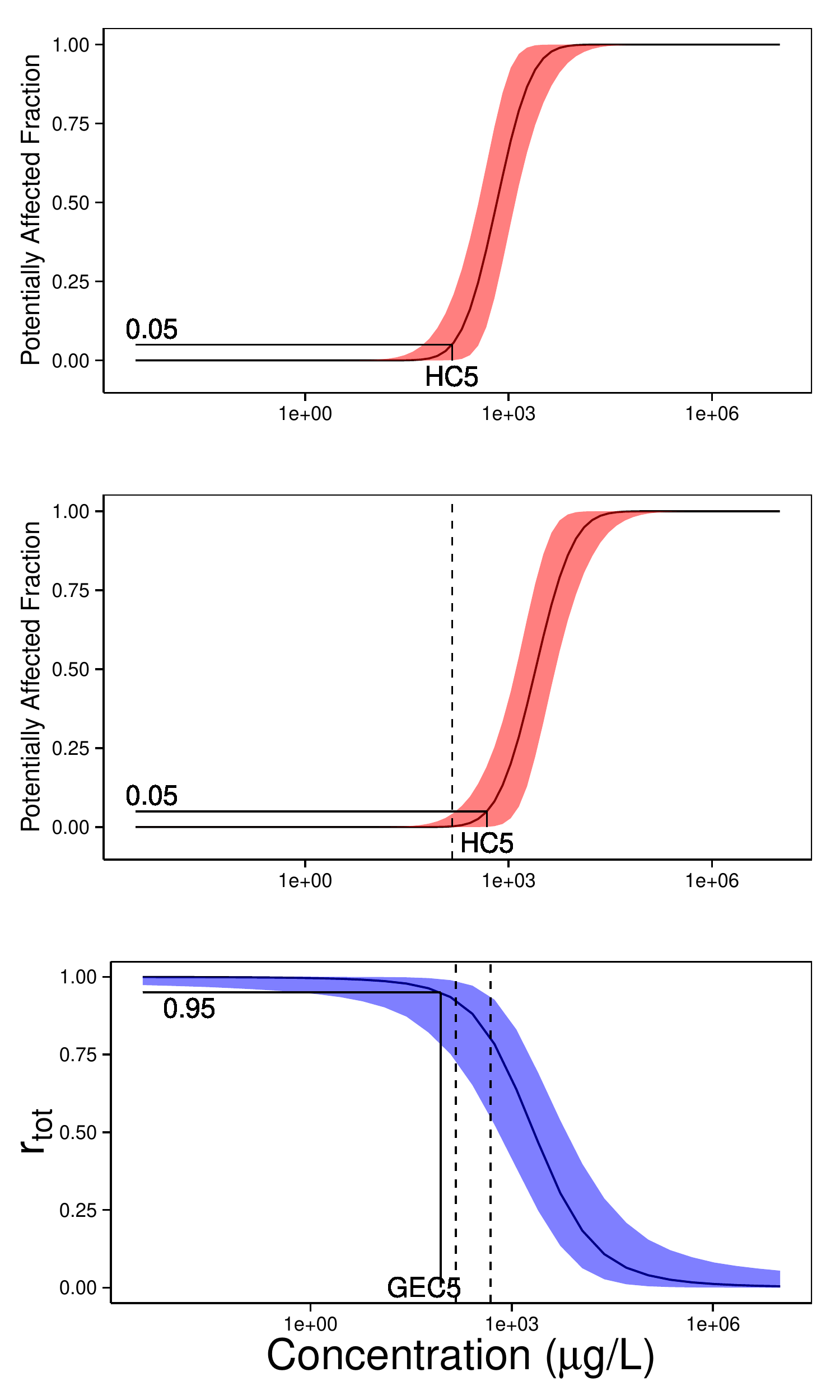}
\includegraphics[width=7.5cm,height=15cm]{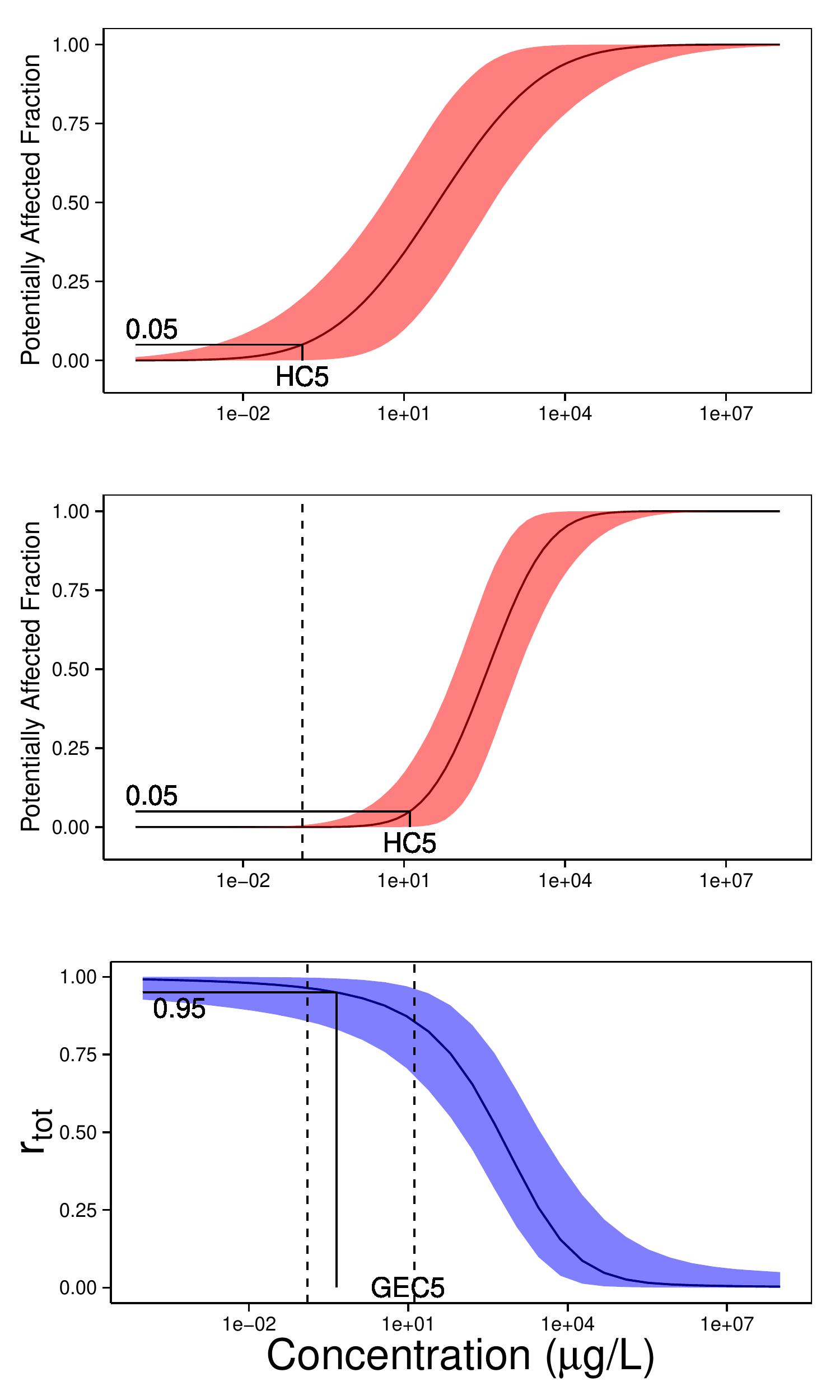}

\captionof{figure}{Species sensitivity distribution and global response of the community for atrazine (left) and diuron (right). Top: classical SSD, built on the $\mathrm{EC}_{10}$ with 95\% bootstrap confidence intervals and the $\mathrm{HC}_5$. Middle: classical SSD, built on the $\mathrm{EC}_{50}$ with bootstrap confidence intervals and the $\mathrm{HC}_5$. Bottom: global response of the community with  95\% credible confidence intervals and the concentration corresponding to a reduction of $5\%$ of the global response ($\mathrm{GEC}_ 5$). The horizontal dotted lines provide visual cues to compare the $\mathrm{HC_{5,EC_{10}}}$, $\mathrm{HC_{5,EC_{50}}}$ and the $\mathrm{GEC_{5}}$.\label{HC_GEC_Diu}}


%
%

\subsection{SSD as a function of the level of effect ($x$ of the $\mathrm{EC}_{x}$)}

Fig. \ref{hc5_vs_x} shows the $\mathrm{HC}_5$ of the diatom community exposed to Diuron, computed from the hierarchical SSD. This is an $\mathrm{HC}_5$ which includes the uncertainty from the raw data. 
The prediction from the hierarchical model was compared to the predictions from the classical SSD. 
The first striking observation is that the classical $\mathrm{HC}_5$ based on the $\mathrm{EC}_{10}$, which ignores the uncertainty from the determination of the CECs, is much higher than the hierarchical $\mathrm{HC}_5$. The second observation is that for a hierarchical $\mathrm{HC}_5$ which includes the original uncertainty on the CECs, the confidence intervals expand wildly for an $x$ of the $\mathrm{EC}_{x}$ below $50$. This can be linked to the fact that for small values of $x$, the uncertainty on the $\mathrm{EC}_{x}$ estimated from a concentration-effect curve is larger than on the $\mathrm{EC}_{50}$ (as was observed on the $\mathrm{EC}_{10}$ on Fig. \ref{fitted_param}). Therefore, in estimating the $\mathrm{HC}_{5}$, the effect of discarding the uncertainty should be greater. 
This phenomenon cannot be observed with classical SSD, since it does not take into account uncertainty on the CECs.
Such an observation contrasts with the reasoning from Aldenberg and Rorije\cite{Aldenberg2013}, which state that taking uncertainty on the CECs into account should increase the value of the $\mathrm{HC}_5$. However valid, their argument cannot be directly applied to our case, for it rests strongly on the assumption of lognormality of the CECs, be they NOEC, QSAR or $\mathrm{EC}_{x}$ at any level of effect. In our model, the $\mathrm{EC}_{50}$ is assumed to follow a lognormal distribution (parameter $b$ and $e$ follow a lognormal distribution), which implies that $\log\mathrm{EC_x}=\log e + \frac{1}{b}\log\left(\frac{x}{1-x}\right)$ for any other $x$ than 50 does not follow a normal distribution. Therefore, it is not surprising to find a hierarchical $\mathrm{HC}_5$ different from the classical $\mathrm{HC}_5$. 


%
%
\begin{center}

\includegraphics[width=0.4\textwidth]{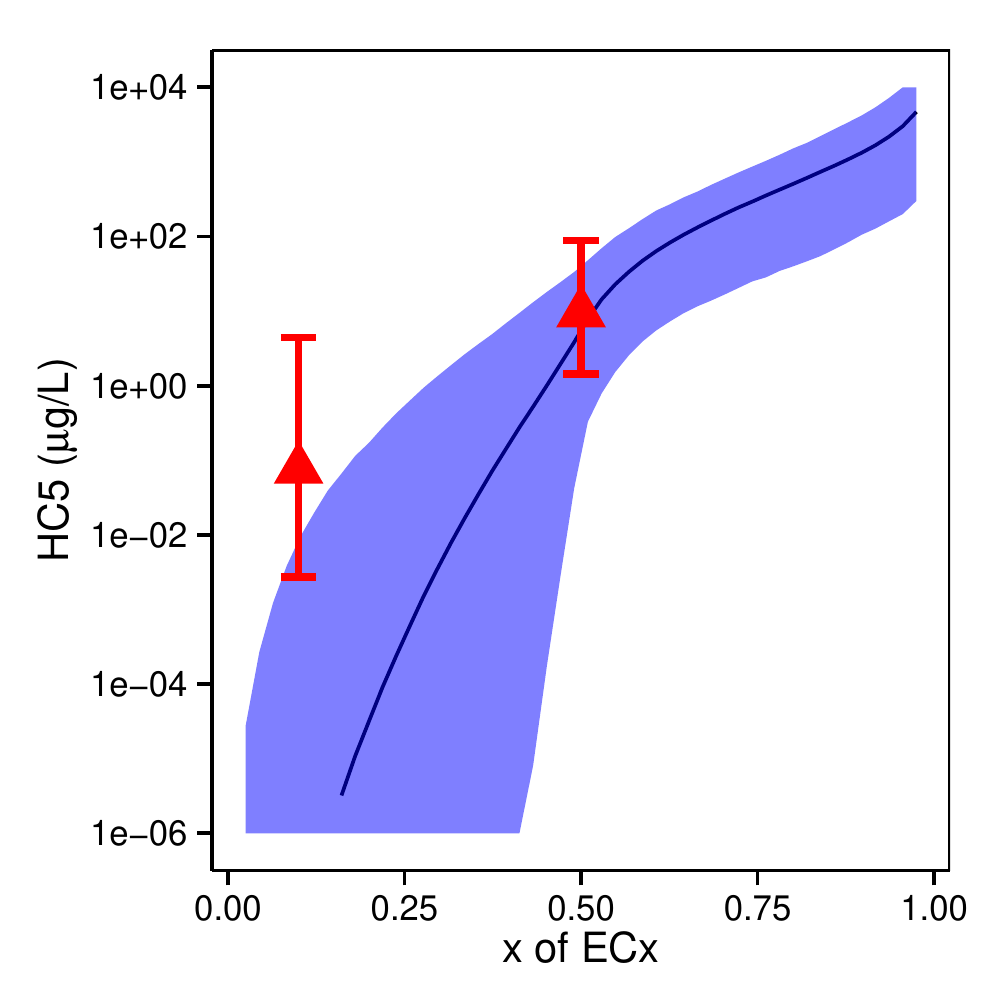}
\captionof{figure}{$\mathrm{HC}_5$ as a function of the $x$ of the $\mathrm{EC}_{x}$ for Diuron obtained from the hierarchical SSD, with the $95\%$ credibility bands. In red, the $\mathrm{HC}_5$ obtained from the classical SSD based on the $\mathrm{EC}_{10}$ and the  $\mathrm{EC}_{50}$ with bootstrap confidence intervals. \label{hc5_vs_x}}
%
%
%

\end{center}
\section{Discussion}


Classical SSDs are widely used to assess risk of chemicals for natural communities, but they present certain limitations\cite{Forbes2002,doi:10.1021/es972418b}. In this study, we presented a hierarchical approach to SSD, which includes all the information present in the raw bioassay data to overcome some of these limitations. This hierarchical SSD differs from classical SSD in that the whole concentration-effect curve is used to build the SSD instead of a single CEC per species. This implies that the hierarchical model requires the full output from bioassays response curves. Unfortunately, such data are not always available. 
For the three parameters loglogistic model used in this study, providing two CECs, such as the $\mathrm{EC}_{10}$ and the $\mathrm{EC}_{50}$ would be sufficient to describe the effect of the contaminant on a species. 
Therefore, reporting only two CECs at the end of a bioassay would be enough to construct a hierarchical SSD in the same spirit as that developed in this work, though without propagating the uncertainty on the CECs. 
Making full use of the bioassay data, the hierarchical SSD propagates not only all the uncertainty from the concentration-effect curve, but also all the information on the shape of the curve. 
It also unveils possible correlations among the parameters, which have a biological significance and might be related to the mode of action of the contaminant. 

One of the advantages of the hierarchical approach is the prediction of the global response of a community presented as a concentration-effect curve which in turn makes it possible to derive a global effect concentration of $x\%$ $\mathrm{GEC}_x$.
 This new kind of threshold does not provide \emph{a priori} information at the species level (what and how much specific species are affected) but it is a tool to make \emph{a priori} risk assessment at the community level (response of all the species together).
 This appeared especially interesting for microbial community, for which chemical effects are often observed and reported at the community level for many endpoints (i.e. respiration, photosynthesis, fluorescence, enzyme activity…). 
This global response does not require the choice of an arbitrary effect level such as the $\mathrm{EC}_{50}$ or $\mathrm{EC}_{10}$.
 Moreover, on the tested contaminants, the hierarchical approach resulted in safe concentration levels which were very close to the classical $\mathrm{HC}_{5}$ defined on the $\mathrm{EC}_{10}$.
 This led us to think that from an operational point of view, the use of the global response should prove as protective as the classical SSD approach. 
The global response may also be used to provide structural or functional information depending on the structural or functional nature of the measured endpoint.
 In that case, the global response would provide information on the functional response of a community and solve one of the problems in the SSD approach \cite{DeLaender2008a,Kefford2012}. Fundamentally, the global response is an indicator containing a radically different type of information compared to SSD. The $\mathrm{HC}_5$ aims to protect 95\% of the species in a community, but there is considerable uncertainty about the fate of the community if the 5\% affected play a key role for some other properties of the community (such as the global response). The $\mathrm{GEC}_{5}$ protects 95\% of the global response, but does not say what proportion of the species are significantly affected (above a given level of effect).
 Together, both SSD and global response provide complementary means to assess the effect of a contaminant on a community.
 Both need to be considered when defining acceptable levels of concentration for a contaminant. 

The definition of the global response strongly depends on the assumption of equipartition of species contribution to the global response. In communities of diatoms, one or several species may dominate and their contribution to the global fluorescence could be preponderant. However, it has been observed that the dominance and the diversity of diatom species within a community change across the seasons\cite{singh2010seasonal,pesce2009response}. Therefore, when considering the fluorescence over a year, the contribution of many species might be averaged, rendering our assumption of equirepartition more plausible. At any rate, this assumption is already present in the classical SSD approach\cite{Forbes2002}. As the simulated species are unidentified, it is not possible to attribute a weight to each of them to sum their fluorescence. 
To circumvent this assumption, it could be possible to define groups of species having comparable fluorescence and define weights according to these groups. 
On a larger dataset, it would certainly be interesting to adopt this approach. 

The first advantage of the hierarchical approach was to introduce more ecological relevance to the risk assessment than the bare classical SSD. The hierarchical approach also provides a perspective on the treatment of uncertainty in the classical SSD. Classical SSD adopts the same approach whatever the level of effect chosen. 
Yet, the degree of uncertainty can strongly depend on the level of effect, and neglecting that uncertainty might certainly bias the estimation of the $\mathrm{HC}_5$. 
In particular, the hierarchical SSD shows that building an SSD on $\mathrm{EC}_{10}$ without considering the uncertainty on these $\mathrm{EC}_{10}$ might lead to a wrong estimate of the $\mathrm{HC}_5$ and of its confidence intervals. 
The hierarchical SSD, which correctly propagates the uncertainty from the raw data to the $\mathrm{HC}_5$ and builds the SSD on any $\mathrm{EC}_{x}$, does not rest on this assumption however. 
We simply assumed that parameter $e$ followed the usual lognormal distribution\cite{Wheeler2002} and opted for the same distribution law for the second parameter. 
With only ten species, there is not much ground to argue for other distributions, but in the future it would be very interesting to analyse larger datasets. 
More tested species would provide a better characterisation of the distribution laws for the concentration-effect model parameters and might support the current distribution choice or guide towards a different structure for the hierarchical model. 
We noted that our result on the hierarchical SSD including all sources of uncertainty and variability contrasted with the argument put forward by Aldenberg and Rorije in \cite{Aldenberg2013}, where they explained that taking uncertainty into account should increase the estimate of the $\mathrm{HC}_5$ compared to a classical SSD approach. We gave a first reason why their argument was not in contradiction with our work: in our model the $\mathrm{EC}_x$ for $x$ different from $50$ do not follow a lognormal distribution. 
A second reason is that in the  hierarchical approach in \cite{Aldenberg2013}, it is assumed that the uncertainty on the CEC (the length of the $95\%$ confidence interval on the CEC) is identical for all species, whereas in our model the uncertainty on the parameters of the concentration-effect model is specific to each species. The species-specific uncertainty mostly depends on the quality of the raw data for that species. 
This is important because the reasoning of Aldenberg and Rorije focuses on the estimation on the variance of the SSD, while including varying degrees of uncertainty for each species could affect the estimation of the mean of the SSD. To understand the role of varying levels of uncertainty, let us consider an extreme case: if there is a large uncertainty on the most sensitive species and a rather small uncertainty on all the other species, we can expect that taking uncertainty into account will shift the estimate of the mean upwards. Since the value of the $\mathrm{HC}_5$ is a function of both the mean and the variance of the SSD, taking uncertainty into account, although reducing the variance of the SSD, does not necessarily increase the estimate of the $\mathrm{HC}_5$.
A third reason stems from the hierarchical structure of the model and the fact that the fit of the concentration effect models at the level of the species is performed in one stroke. The fit of the $\mathrm{b}$ and $\mathrm{e}$ parameters for one species is influenced by the data from the other species. More specifically, since the tested species are assumed to come from the same community with the same species sensitivity distribution, the estimation of the concentration-effect model parameters is the result of information coming from all species together. On the contrary, in classical SSD the fit of the concentration effect parameters obtained by linear regression depends solely on the data for one species. Therefore, the $e$ parameter for a given species estimated in the hierarchical model can be slightly different from the $e$ parameter estimated by nonlinear regression. Translated at the community level, this implies that the value of the $\mathrm{HC}_5$ is not determined by the value of the CECs and their uncertainty, but by a more subtle interplay between the raw data and the distribution law of species sensitivity in the community. For all these reasons, we do not believe that our results are incompatible with previous work by Aldenberg and Rorije.

As a conclusion, the current hierarchical modelling of SSD aimed to include all the experimental data into the SSD. The first step was to avoid summarizing the full concentration-effect curve by a single critical effect concentration. However, only data at the end of the experiment were used. Bioassay data often include a tracking over time of the contaminant effect and this information could be included as well in the SSD. 
Modelling time-dependence would essentially consists in adding a supplementary level to the hierarchy. 
Studying the time component of  SSD is particularly interesting  because toxicity of a contaminant clearly evolves over time, yet the observation period is often constrained by practical considerations\cite{Fox2013}. Future work will focus on including time dependence into the SSD approach to improve the accuracy and the biological relevance of its predictions. 


\section*{Acknowledgements}
Funding for this project was provided by  a grant from la Région Rhône-Alpes.

\bibliography{library}

\begin{thebibliography}{10}

\bibitem{Aldenberg2000a}
T.~Aldenberg and J.~Jaworska.
\newblock {Uncertainty of the hazardous concentration and fraction affected for
  normal species sensitivity distributions.}
\newblock {\em Ecotoxicology and Environmental Safety}, 46(1):1--18, May 2000.

\bibitem{Aldenberg2013}
T.~Aldenberg and E.~Rorije.
\newblock {Species Sensitivity Distribution estimation from uncertain
  (QSAR-based) effects data.}
\newblock {\em Alternatives to Laboratory Animals}, 41(1):19--31, Mar. 2013.

\bibitem{Aldenberg1993}
T.~Aldenberg and W.~Slob.
\newblock {Confidence limits for hazardous concentrations based on logistically
  distributed NOEC toxicity data}.
\newblock {\em Ecotoxicology and Environmental Safety}, 25(1):48--63, 1993.

\bibitem{Verdonck2001}
F.~{A.M Verdonck}, J.~Jaworska, O.~Thas, P.~A. Vanrolleghem, and F.~a.M
  Verdonck.
\newblock {Determining environmental standards using bootstrapping, bayesian
  and maximum likelihood techniques: a comparative study}.
\newblock {\em Analytica Chimica Acta}, 446(1-2):427--436, Nov. 2001.

\bibitem{Baty2013}
F.~Baty and M.~L. Delignette-Muller.
\newblock {\em nlstools: tools for nonlinear regression diagnostics}, 2013.

\bibitem{Brooks1998}
S.~P. Brooks and A.~Gelman.
\newblock {General methods for monitoring convergence of iterative
  simulations}.
\newblock {\em Journal of computational and graphical statistics},
  7(4):434--455, 1998.

\bibitem{DeLaender2008a}
F.~{De Laender}, K.~a.~C. {De Schamphelaere}, P.~A. Vanrolleghem, and C.~R.
  Janssen.
\newblock {Is ecosystem structure the target of concern in ecological effect
  assessments?}
\newblock {\em Water research}, 42(10-11):2395--402, May 2008.

\bibitem{Dowse2013}
R.~Dowse, D.~Tang, C.~G. Palmer, and B.~J. Kefford.
\newblock {Risk assessment using the species sensitivity distribution method:
  data quality versus data quantity.}
\newblock {\em Environmental Toxicology and Chemistry}, 32(6):1360--1369, June
  2013.

\bibitem{efron1993introduction}
B.~Efron and R.~Tibshirani.
\newblock {\em {An introduction to the bootstrap}}, volume~57.
\newblock CRC press, 1993.

\bibitem{Forbes2002}
V.~E. Forbes and P.~Calow.
\newblock {Species sensitivity distributions revisited: a critical appraisal}.
\newblock {\em Human and Ecological Risk Assessment}, 8(3):473--492, 2002.

\bibitem{Fox2013}
D.~R. Fox and E.~Billoir.
\newblock {Time-dependent species sensitivity distributions.}
\newblock {\em Environmental toxicology and chemistry / SETAC}, 32(2):378--83,
  Feb. 2013.

\bibitem{Grist2009a}
E.~P.~M. Grist and K.~M.~Y. Leung.
\newblock {Better bootstrap estimation of hazardous concentration thresholds
  for aquatic assemblages}.
\newblock {\em Environmental Toxicology and Chemistry}, 21(7):1515--1524, 2009.

\bibitem{Jagoe1997}
R.~H. Jagoe and M.~C. Newman.
\newblock {Bootstrap estimation of community NOEC values}.
\newblock {\em Ecotoxicology}, 6(5):293--306, 1997.

\bibitem{Kefford2012}
B.~J. Kefford, R.~B. Sch\"{a}fer, and L.~Metzeling.
\newblock {Risk assessment of salinity and turbidity in Victoria (Australia) to
  stream insects' community structure does not always protect functional
  traits.}
\newblock {\em The Science of the total environment}, 415:61--8, Jan. 2012.

\bibitem{ETC:ETC2644}
G.~{Kon Kam King}, P.~Veber, S.~Charles, and M.~L. Delignette-Muller.
\newblock {MOSAIC-SSD: A new web-tool for species sensitivity distribution to
  include censored data by maximum likelihood}.
\newblock {\em Environmental Toxicology and Chemistry}, pages n/a----n/a, 2014.

\bibitem{Larras2012}
F.~Larras, A.~Bouchez, F.~Rimet, and B.~Montuelle.
\newblock {Using Bioassays and Species Sensitivity Distributions to Assess
  Herbicide Toxicity towards Benthic Diatoms.}
\newblock {\em PloS one}, 7(8):e44458, Jan. 2012.

\bibitem{Moore2010}
D.~Moore, W.~Warren-Hicks, S.~Qian, A.~Fairbrother, T.~Aldenberg, T.~Barry,
  R.~Luttik, and H.~Ratte.
\newblock {Uncertainty analysis using classical and bayesian hierarchical
  models}.
\newblock In W.~Warren-Hicks and A.~Hart, editors, {\em Application of
  Uncertainty Analysis to Ecological Risk of Pesticides}, pages 134--141. CRC
  press, Pensacola (Florida), US, 2010.

\bibitem{pesce2009response}
S.~Pesce, I.~Batisson, C.~Bardot, C.~Fajon, C.~Portelli, B.~Montuelle, and
  J.~Bohatier.
\newblock {Response of spring and summer riverine microbial communities
  following glyphosate exposure}.
\newblock {\em Ecotoxicology and environmental safety}, 72(7):1905--1912, 2009.

\bibitem{plummer2003jags}
M.~Plummer.
\newblock {JAGS: A program for analysis of Bayesian graphical models using
  Gibbs sampling}.
\newblock In {\em Proceedings of the 3rd International Workshop on Distributed
  Statistical Computing (DSC 2003). March}, pages 20--22, 2003.

\bibitem{Posthuma2010}
L.~Posthuma, G.~W. {Suter II}, and T.~P. Traas.
\newblock {\em {Species sensitivity distributions in ecotoxicology}}.
\newblock CRC press, 2010.

\bibitem{doi:10.1021/es972418b}
M.~Power and L.~S. McCarty.
\newblock {Fallacies in Ecological Risk Assessment Practices}.
\newblock {\em Environmental Science and Technology}, 31(8):370A----375A, 1997.

\bibitem{Shao2000}
Q.~Shao.
\newblock {Estimation for hazardous concentrations based on NOEC toxicity data:
  an alternative approach}.
\newblock {\em Environmetrics}, 11(5):583--595, Sept. 2000.

\bibitem{singh2010seasonal}
M.~Singh, P.~Lodha, and G.~P. Singh.
\newblock {Seasonal diatom variations with reference to physico-chemical
  properties of water of Mansagar lake of Jaipur, Rajasthan}.
\newblock {\em Research Journal of Agricultural Sciences}, 1(4):451--457, 2010.

\bibitem{VanderHoeven2001}
N.~van~der Hoeven.
\newblock {Estimating the 5-percentile of the species sensitivity distributions
  without any assumptions about the distribution.}
\newblock {\em Ecotoxicology}, 10(1):25--34, Feb. 2001.

\bibitem{Wagner1991}
C.~Wagner and H.~Lokke.
\newblock {Estimation of ecotoxicological protection levels from NOEC toxicity
  data}.
\newblock {\em Water Research}, 25(10):1237--1242, 1991.

\bibitem{weisberg2005applied}
S.~Weisberg.
\newblock {\em {Applied linear regression}}, volume 528.
\newblock John Wiley $\backslash$\& Sons, 2005.

\bibitem{Wheeler2002}
J.~R. Wheeler, E.~P.~M. Grist, K.~M.~Y. Leung, D.~Morritt, and M.~Crane.
\newblock {Species sensitivity distributions: data and model choice.}
\newblock {\em Marine Pollution Bulletin}, 45(1-12):192--202, Jan. 2002.

\end{thebibliography}
\bibliographystyle{abbrv}

\section*{Supplementary information}

\subsection*{A) Fit at the species level for the other herbicides}


\begin{minipage}[c]{1.05\linewidth}
\includegraphics[width=0.4\textwidth]{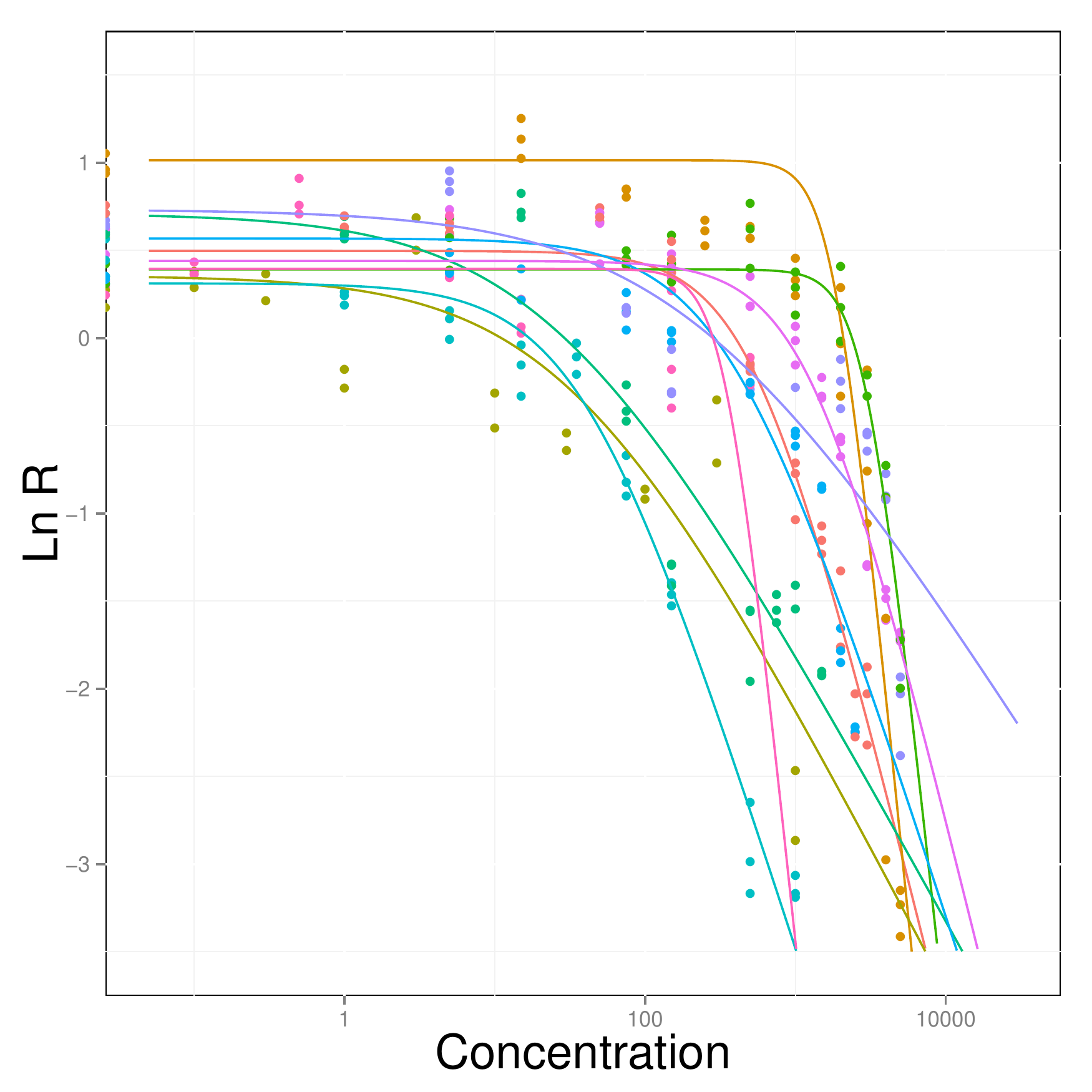}
\includegraphics[width=0.4\textwidth]{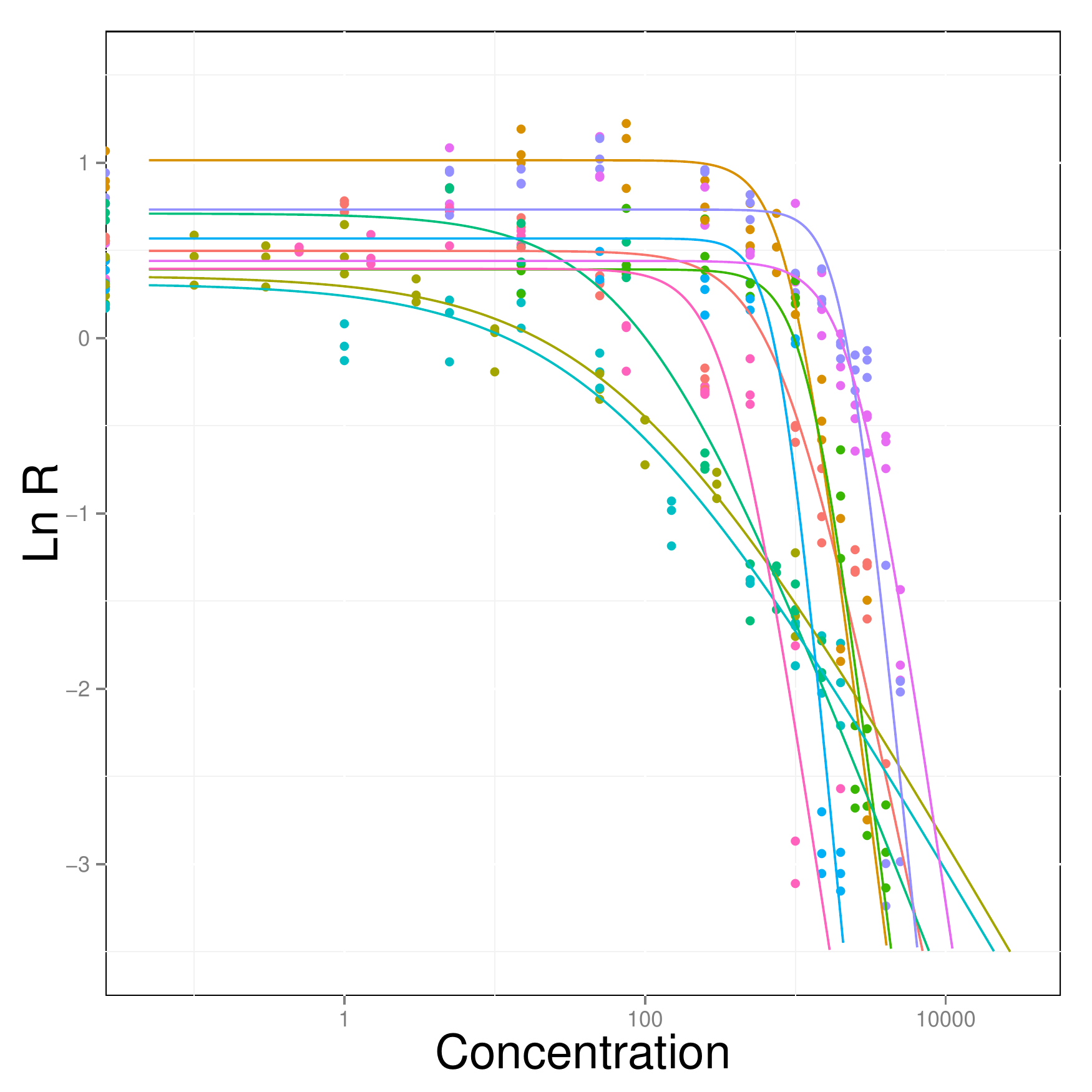}\\

\includegraphics[width=0.4\textwidth]{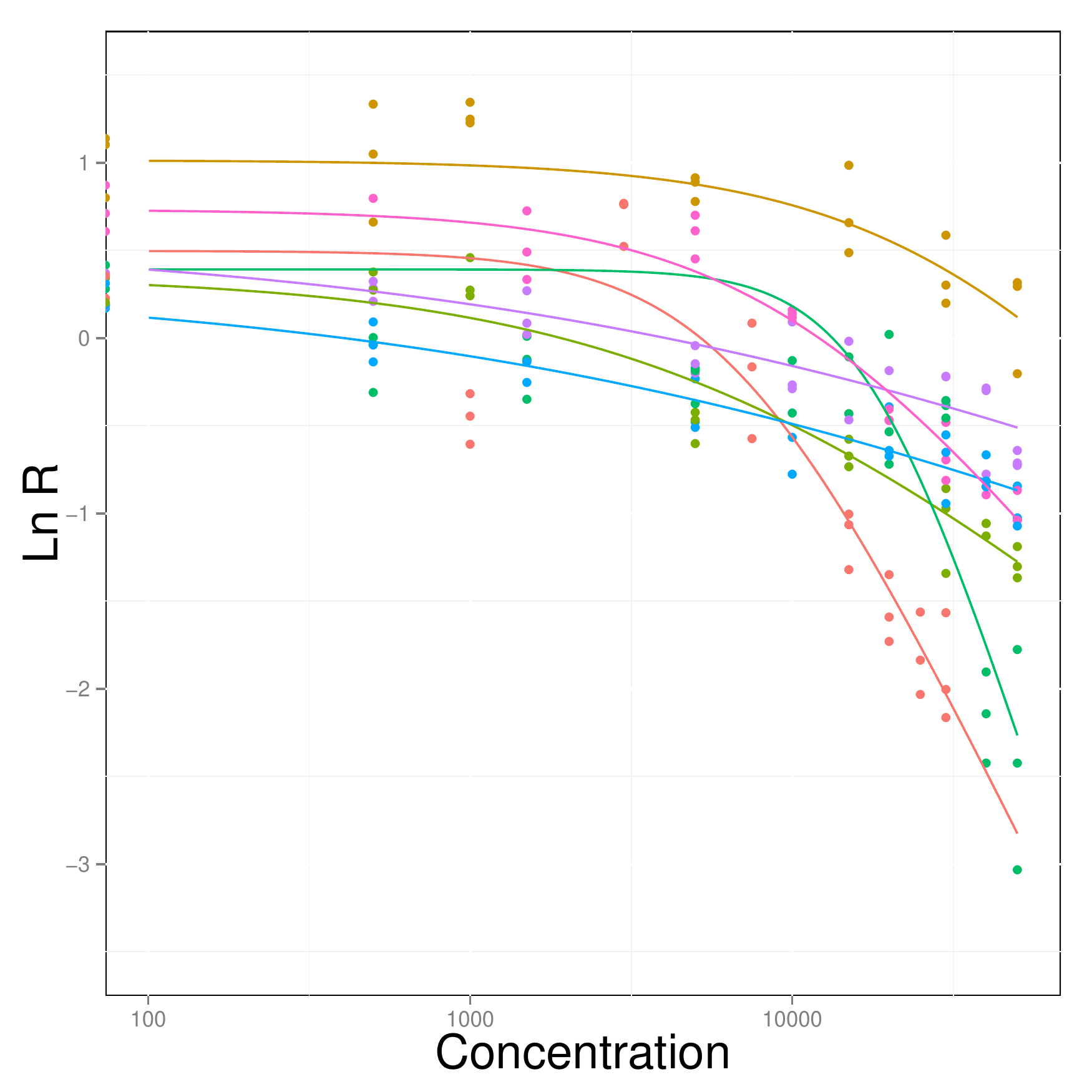}
\includegraphics[width=0.4\textwidth]{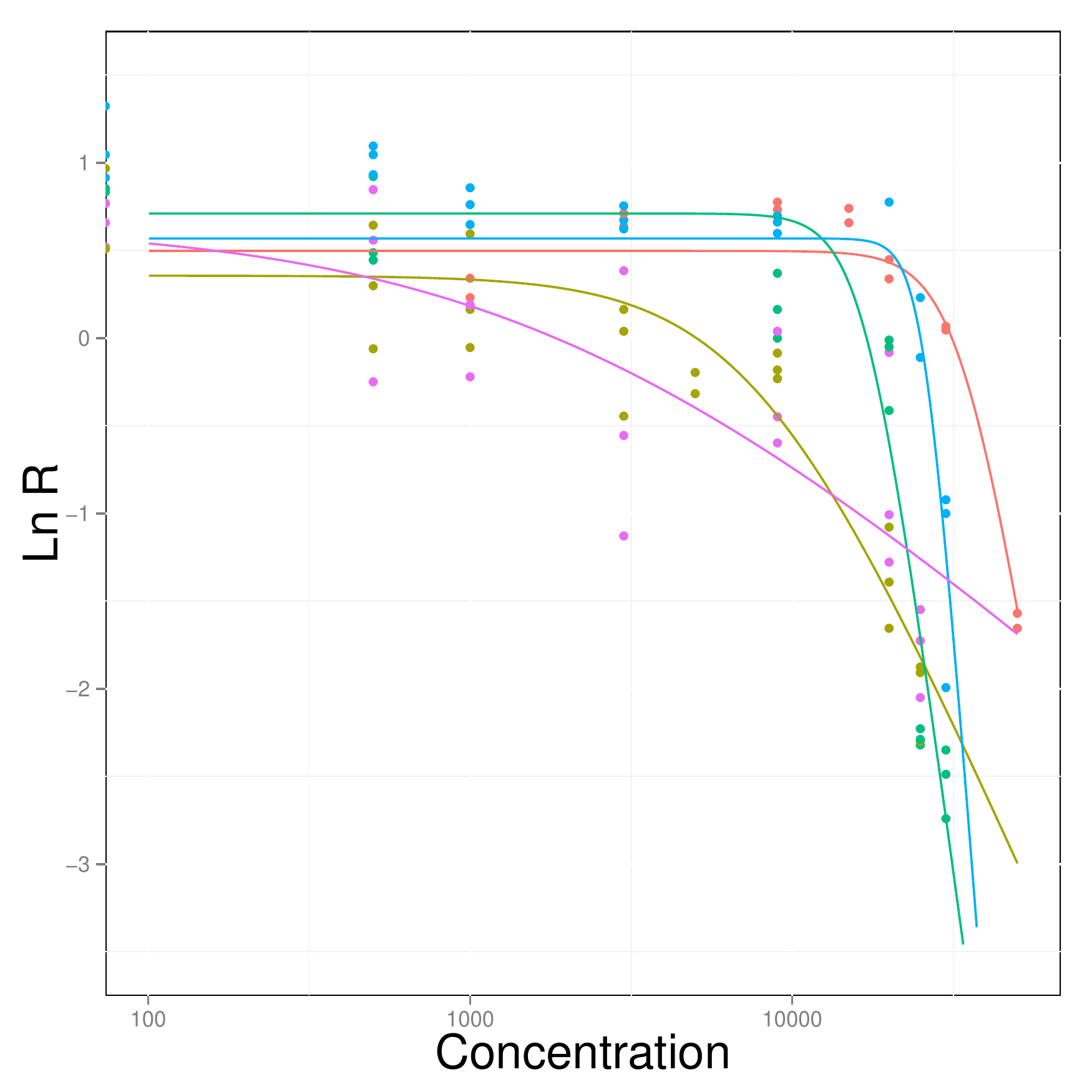}
\end{minipage}

\captionof{figure}{Original data and fit of the model at the level of the original diatom species for terbutryne (top left), isoproturon (top right), metolachlor (bottom left) and Dimetachlor (bottom right). Each colour denotes a different species. The rectangle denotes the control response in log scale. Parameters for the curves are the median values of the marginal parameter distributions.\label{fit}}

\subsection*{B) R script to fit the hierarchical model}

To run the example script, you must:

\begin{enumerate}
\item install R (http://cran.r-project.org/)
\item install JAGS (http://mcmc-jags.sourceforge.net/)
\item install the R package rjags within R 
\item It is recommanded but not compulsory to install the R package dclone
\item run the script written in file run\_mcmc.R, for instance using the command Rscript run\_mcmc.R
\end{enumerate}

File run\_mcmc.R contains an R function to calculate posterior distribution of the parameters of the hierarchical model. It also contains a computation of the Gelman and Rubin diagnostic, and a traceplot of the posterior distributions.

If there are several cores on the computer, it is possible to run the MCMC chains in parallel. To run the mcmc algorithm in parallel, comment out the sequential version and uncomment the parallel version.

File data.db contains the Diuron dataset, file model.txt contains the jags model.
%
%
%
%
\end{document}